\documentclass[aip,cha,reprint]{revtex4-1}

\usepackage{color}
\usepackage{graphicx} % inclusion de graficos
\usepackage{xfrac} %to use \sfrac
\usepackage{amssymb, amsmath, amsthm, amsfonts}
\newcommand{\BE}{\begin{equation}}
\newcommand{\EE}{\end{equation}}
\newcommand{\BA}{\begin{eqnarray}}
\newcommand{\EA}{\end{eqnarray}}
\newcommand{\bx}{{\bf x}}

\newcommand{\degree}{^\circ}

\begin{document}

% Use the \preprint command to place your local institutional report number
% on the title page in preprint mode.
% Multiple \preprint commands are allowed.
%\preprint{}

\title{Dominant transport pathways in an atmospheric blocking event} %Title of paper
%\title{Flow networks: Networks of Lagrangian transport}

% repeat the \author .. \affiliation  etc. as needed
% \email, \thanks, \homepage, \altaffiliation all apply to the current author.
% Explanatory text should go in the []'s,
% actual e-mail address or url should go in the {}'s for \email and \homepage.
% Please use the appropriate macro for the type of information

% \affiliation command applies to all authors since the last \affiliation command.
% The \affiliation command should follow the other information.

\author{Enrico Ser-Giacomi}
\affiliation{IFISC (CSIC-UIB), Instituto de F\'{\i}sica
Interdisciplinar y Sistemas Complejos, Campus Universitat de
les Illes Balears, E-07122 Palma de Mallorca, Spain}

\author{Ruggero Vasile}
\affiliation{Ambrosys GmbH,
Albert-Einstein-Str. 1-5 , 14473 Potsdam, Germany}

\author{Irene Recuerda}
\affiliation{IFISC (CSIC-UIB), Instituto de F\'{\i}sica
Interdisciplinar y Sistemas Complejos, Campus Universitat de
les Illes Balears, E-07122 Palma de Mallorca, Spain}

\author{Emilio Hern\'{a}ndez-Garc\'{\i}a}
\affiliation{IFISC (CSIC-UIB), Instituto de F\'{\i}sica
Interdisciplinar y Sistemas Complejos, Campus Universitat de
les Illes Balears, E-07122 Palma de Mallorca, Spain}

\author{Crist\'obal L\'{o}pez}
\affiliation{IFISC (CSIC-UIB), Instituto de F\'{\i}sica
Interdisciplinar y Sistemas Complejos, Campus Universitat de
les Illes Balears, E-07122 Palma de Mallorca, Spain}

%\date{\today}
%\date{January 15, 2015}  % put the final date
\date{First version, January 15 2015; this version, July 28 2015}

% Collaboration name, if desired (requires use of superscriptaddress option in \documentclass).
% \noaffiliation is required (may also be used with the \author command).
%\collaboration{}
%\noaffiliation

\date{\today}
%\date{September 5, 2014}   %put the final date

\begin{abstract}
A Lagrangian flow network is constructed for the atmospheric
blocking of eastern Europe and western Russia in summer 2010.
We compute the most probable paths followed by fluid particles
which reveal the {\it Omega}-block skeleton of the event. A
hierarchy of sets of highly probable paths is introduced to
describe transport pathways when the most probable path alone
is not representative enough. These sets of paths have the
shape of narrow coherent tubes flowing close to the most
probable one. Thus, even when the most probable path is not
very significant in terms of its probability, it still
identifies the geometry of the transport pathways.
\end{abstract}

%\pacs{}% insert suggested PACS numbers in braces on next line

\maketitle %\maketitle must follow title, authors, abstract and \pacs

% Body of paper goes here. Use proper sectioning commands.
% References should be done using the \cite, \ref, and \label commands

\begin{quotation}
Eastern Europe and Western Russia experienced a strong heat
wave with devastating consequences in the summer of 2010. This
was due to an atmospheric blocking episode that lasted during
several weeks. Despite these type of events have been
well-investigated over the years, a complete understanding and
prediction is still missing. In this work we present a
characterization of this flow pattern based on the study of
fluid transport as a Lagrangian flow network, so that the
methodology of complex networks can be applied. In particular,
the most probable paths linking nodes of this atmospheric
network reveal the dominant pathways traced by atmospheric
fluid particles.
\end{quotation}

% If in two-column mode, this environment will change to single-column format so that long equations can be displayed.
% Use only when necessary.
%\begin{widetext}
%$$\mbox{put long equation here}$$
%\end{widetext}

% Figures should be put into the text as floats.
% Use the graphics or graphicx packages (distributed with LaTeX2e).
% See the LaTeX Graphics Companion by Michel Goosens, Sebastian Rahtz, and Frank Mittelbach for examples.
%
% Here is an example of the general form of a figure:
% Fill in the caption in the braces of the \caption{} command.
% Put the label that you will use with \ref{} command in the braces of the \label{} command.
%
% \begin{figure}
% \includegraphics[width=0.5\columnwidth]{}%
% \caption{\label{}}%
% \end{figure}

% Tables may be be put in the text as floats.
% Here is an example of the general form of a table:
% Fill in the caption in the braces of the \caption{} command. Put the label
% that you will use with \ref{} command in the braces of the \label{} command.
% Insert the column specifiers (l, r, c, d, etc.) in the empty braces of the
% \begin{tabular}{} command.
%
% \begin{table}
% \caption{\label{} }
% \begin{tabular}{}
% \end{tabular}
% \end{table}

\section{Introduction}
\label{sec:intro}

Lagrangian analysis of transport in fluids, in particular in
geophysical and time-dependent contexts, has experienced
intense developments in the last decades. These can be roughly
classified in three classes: Some of the approaches search for
geometric objects --lines, surfaces, usually related to
invariant manifolds -- which bound fluid regions with different
properties
\cite{mancho2006tutorial,haller2012geodesic,balasuriya2012explicit}.
In the second type of approaches one computes different types
of Lyapunov exponents and other stretching-like fields in the
fluid domain
\cite{haller2001distinguished,joseph2002relation,dovidio2004mixing,mancho2013lagrangian}.
Finally, set-oriented methods
\cite{froyland2003detecting,dellnitz2009seasonal,
froyland2010transport,levnajic2010ergodic,froyland2012three}
address directly the motions of finite-size regions.

Most of these techniques focus in identifying proper
\emph{Lagrangian Coherent Structures}
\cite{haller2000lagrangian,peacock2010introduction,haller2015lagrangian},
understood as \emph{barriers to transport} or \emph{coherent
regions} with small fluid exchange with the surrounding medium.
Much less is known about the actual \emph{routes of transport},
the dominant \emph{pathways} along which fluid particles travel
and fluid properties are interchanged.

In principle, the pathways are simply given by trajectories
starting from the desired initial conditions. This is true when
the advection dynamics is represented by a deterministic
dynamical system and the initial condition is precisely fixed.
In many applications however, particularly in geosciences,
stochastic components are added to the motions to better
represent unresolved spatial scales
\cite{thomson1987criteria,stohl1998computation,veneziani2004oceanic}.
Also, imprecisely stated initial conditions will develop into a
divergent set of possible trajectories, because of the
inherently chaotic character of advection by nearly any
nontrivial fluid flow, particularly when it is time-dependent.
In fact in real experiments such as in the deployment of buoys
or balloons the trajectories of closely released objects
diverge soon
\cite{trounday1995dispersion,lacasce2008statistics,lumpkin2010surface}.
The so-called \emph{spaghetti plots}
\cite{veneziani2004oceanic} provide a visual representation of
this dispersion. But they become, when many trajectories are
represented, cluttered and unclear. Some type of clustering or
the selection of relevant trajectories is needed to highlight
which are the dominant routes among a large set of possible
trajectories.

We have recently developed \cite{sergiacomi2015most} a
formalism that computes, in unsteady flows, the optimal fluid
paths starting at given initial conditions and also optimal
paths connecting pairs of points. By \emph{optimal} we refer to
the paths which are more likely to be followed, in a
well-defined sense made explicit below, by the fluid particles
initialized in a finite neighborhood of the initial locations.
By this reason they are called \emph{most probable paths}. The
methodology builds on the set-oriented techniques
\cite{froyland2003detecting,dellnitz2009seasonal,
froyland2010transport,levnajic2010ergodic,froyland2012three}
which discretize space to provide a coarse-grained description
of transport, and draws analogies with network theory
\cite{dellnitz2006graph,santitissadeekorn2007identifying,rossi2014hydrodynamic,sergiacomi2015flow,sergiacomi2015most},
for which tools to compute optimal paths in graphs are well
developed. A related formalism addressing optimal paths in
time-independent flows in continuous time has been developed by
Metzner et al.\cite{metzner2009transition}. The optimal paths
provide the main pathways or skeleton of the transport process
in a given geographical area. Because of the implicit
stochastic ingredient in the coarse-graining procedure of
set-oriented methods, this methodology, at variance with other
ones more tied to the theory of smooth dynamical systems, can
be applied equally well to cases of deterministic transport and
to strongly diffusive situations.

In this paper we compute optimal transport paths for the
atmospheric circulation during a blocking event occurring in
Summer 2010 (in particular we focus our study for the period
20th July - 30th July) over Eastern Europe and Russia. This
atmospheric flow has very different temporal and spatial
scales, and is much more diffusive, than the oceanic flow
analyzed in Ser-Giacomi at al. \cite{sergiacomi2015most}. We
give a more detailed description of the methodology sketched in
that reference, and generalize it to extend the concept of most
probable path to a hierarchy of sets of paths characterized by
an increasing probability. The spatial coherence of these sets
is also discussed.

The paper is organized as follows: In Sect. \ref{sec:optimal}
we summarize the definition and construction of the optimal
pathways as {\sl most probable paths} in a flow network. In
Sect. \ref{sec:hpps} we extend this concept to {\sl sets} of
highly probable paths and give rules to establish their
significance and spatial coherence. Sect. \ref{sec:atm}
describes the atmospheric blocking event, the data and models
we use to compute the Lagrangian trajectories, and construct
the flow network from them. Sect. \ref{sec:results} contains
our results: optimal pathways for different dates and
locations, and also a discussion of the statistical
representativeness of the optimal paths on the sets of highly
probable paths. The final Section summarizes our Conclusions.
An Appendix applies our formalism to a simple model flow, an
analytic double-gyre system, so that the properties of the
optimal and highly probable paths computed for the atmospheric
dynamics could be more easily understood in this simplified
framework.

\section{Optimal paths from Lagrangian flow networks}
\label{sec:optimal}

Our approach to find optimal paths in time-dependent fluid
flows first represents the fluid transport dynamics as a
time-dependent flow network \cite{sergiacomi2015flow} and then
uses graph-theory techniques to extract from it these optimal
paths. Following the set-oriented methodology
\cite{froyland2003detecting,dellnitz2009seasonal,
froyland2010transport,levnajic2010ergodic,froyland2012three,sergiacomi2015flow}
we proceed first by a discretization of the spatial domain of
interest, dividing it into $N$ non-overlapping boxes. In terms
of the network-theory approach to
transport\cite{dellnitz2006graph,sergiacomi2015flow,sergiacomi2015most}
each of these boxes will represent a single network
\emph{node}. A large number of ideal fluid particles is
released in each box. Under advection by a given velocity
field, links between nodes are established by studying the
Lagrangian trajectories of the particles exchanged among each
pair of network nodes. This is conveniently done with a
temporal discretization, i.e. we consider the dynamics
restricted to a time interval $[t_0,t_M]$ and divide it in time
steps of length $\tau$, $t_l = t_0 +l\tau,\quad l = 0,1...,M$.
For each time interval $[t_{l-1},t_l]$ we integrate the
equations of motion of each ideal fluid particle and keep track
of each trajectory. The transport dynamics will then be
described by adjacency matrices $\mathbf{A}^{(l)},\,(l=1...M)$,
in which a matrix element $\mathbf{A}_{IJ}^{(l)}$ is given by
the number of particles initialized at time $t_{l-1}$ in node
$I$ that end up at time $t_l$ in node $J$. Since the velocity
field will vary in time the adjacency matrices will depend on
the time interval considered. The weighted network we build
will therefore have an explicit time-dependent character and
can be analyzed, for instance, using \emph{time-ordered
graphs}\cite{kim2012temporal,sergiacomi2015most}.

A fundamental assumption we make is that of a Markovian
dynamics, i.e. at each time interval the ideal fluid particles
are initialized with uniform density in each box, thus without
keeping track of the trajectories at the previous time step.
The effect of such assumption is to introduce diffusive effects
in the dynamics even when the original equations of motion are
fully deterministic \cite{froyland2013analytic}. In the limit
of very small boxes and very short time steps, this
computational diffusion is suppressed and we approach the
perfect Lagrangian motion under the given velocity field (which
itself can contain diffusive or fluctuating terms).

In our network approach spatio-temporal particle trajectories
are mapped into discretized paths between the network nodes. We
define an $M$-step path $\mu$ between nodes $I$ and $J$ as the
ordered sequence of $(M+1)$ nodes, $\mu = \{I, k_1,
...,k_{M-1},J\}$, crossed to reach node $J$ at time $t_M$
starting from node $I$ at time $t_0$. Under the Markovian
hypothesis we can associate a probability to each of these
paths as
\begin{equation}\label{probability}
(p^M_{IJ})_{\mu}=\mathbf{T}_{Ik_1}^{(1)}\biggl[\prod_{l=2}^{M-1}\mathbf{T}_{k_{l-1}k_l}^{(l)}\biggl]\mathbf{T}_{k_{M-1}J}^{(M)},
\end{equation}
where
\begin{equation}
\mathbf{T}_{k_{l-1}k_l}^{(l)}=\frac{\mathbf{A}_{k_{l-1}k_l}^{(l)}}{s_{out}^{(l)}(k_{l-1})}
\end{equation}
is the probability of a fluid particle to reach node $k_l$ at
time $t_l$ if it was initialized at time $t_{l-1}$ in node
$k_{l-1}$, estimated as the ratio of the number of particles
doing so to the total number of particles released at the
initial node and time. The quantity $s_{out}^{(l)}(k) =
\sum_j\mathbf{A}_{kj}^{(l)}$ is called \emph{out-strength} of
node $k$ during the $l$-th time step.

\begin{figure}[!]
\begin{center}
	\includegraphics[width=\columnwidth]{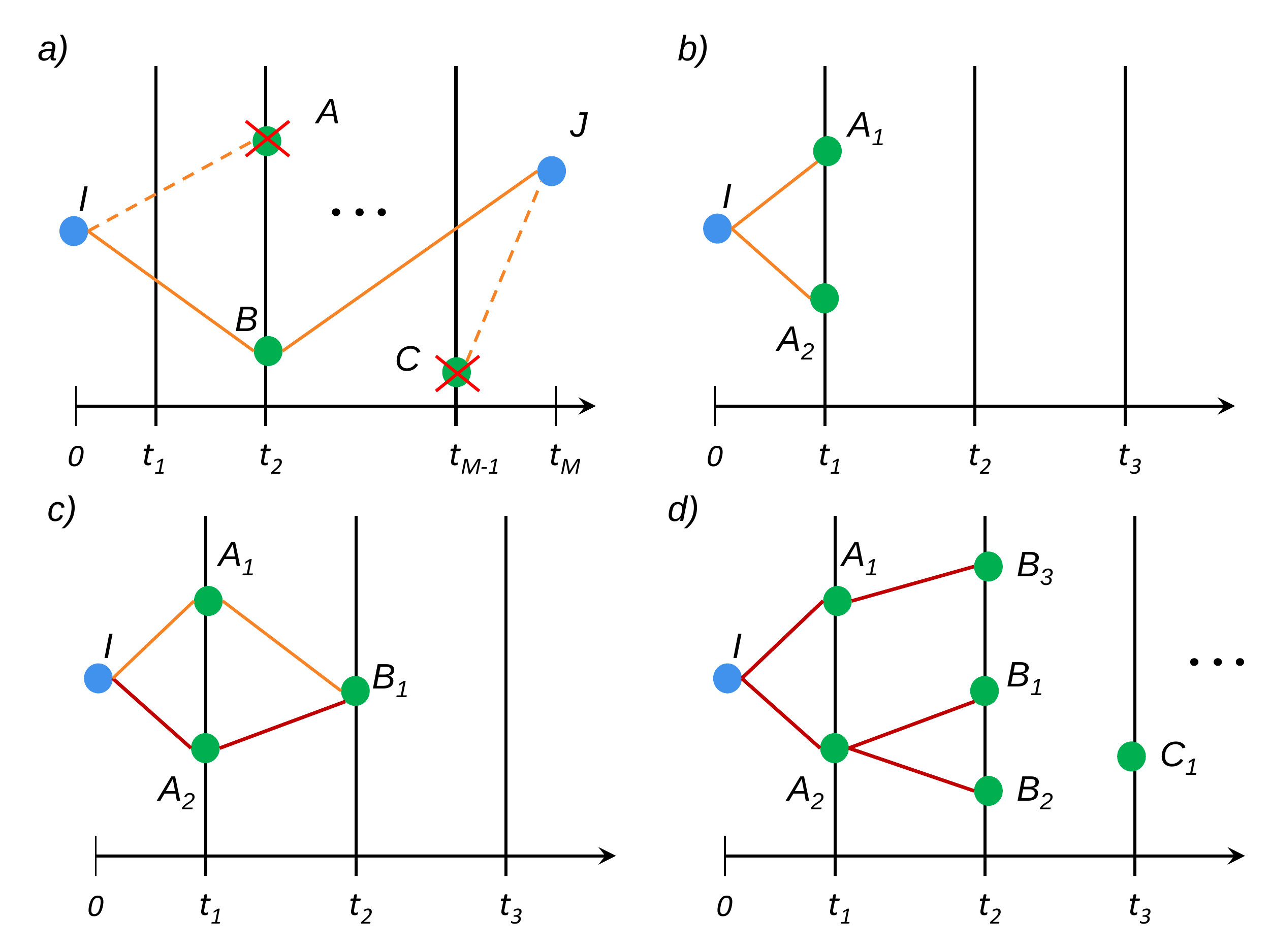}
\end{center}
\caption{Schematics of the algorithm to find the MPP of $M$ steps between $I$ and $J$. a) First part:
determination of the accessible nodes. Point $A$ is reachable from $I$ at
$t = t_2$ but it is not possible to reach $J$ from it in the rest of the time interval. Point $C$ is
not reachable from $I$ at $t = t_{M-1}$ even if $J$ can be reached from it. Point $B$ satisfies
both accessibility conditions, therefore, in contrast to points $A$ and $C$, it belongs to the
accessibility set and it will be considered in the calculation of the MPP.
Systematic identification of all accessible nodes is done my applying the criteria in
Eq. (\ref{accessibility}). The rest of the figure illustrates the recursive
maximization procedure given by Eq. (\ref{recursive}): b) In the first time step one assigns to the
links towards the nodes $A_1$ and $A_2$ (considered to be the only ones in the accessibility
set $\mathbf{U}^{(1)}_{IJ}$) the probabilities $\mathbf{T^{(1)}}_{I A_1}$ and
$\mathbf{T^{(1)}}_{I A_2}$, respectively. c) For node $B_1$ one considers the links
from $A_1$ and $A_2$, evaluates the path's probabilities
$\mathbf{T^{(1)}}_{I A_1} \mathbf{T^{(2)}}_{A_1 B_1}$
and $\mathbf{T^{(1)}}_{I A_2} \mathbf{T^{(2)}}_{A_2 B_1}$, and selects the maximum one
(in the figure the corresponding to the path $I,A_2 B_1$, red lines). One repeats this for all
nodes $B_1,B_2, B_3$ in the accessibility
set $\mathbf{U}^{(2)}_{IJ}$ to obtain the MPPs between $I$ and these nodes, and then
the procedure can be iterated again for the accessible nodes at time $t_3$.}
\label{fig:algorithm}
\end{figure}

Among all possible $M$-step paths between node $I$ and $J$ the
one associated with the highest probability in
Eq.~\eqref{probability} is called the most probable path (MPP)
and is denoted by $\eta_{IJ}^M$. Since this path depends
explicitly on the number $M$ of steps considered, it could be
also named ``fixed-time most probable path". Its probability is
denoted by $P_{IJ}^M = \max_{\mu}\{(p^M_{IJ})_{\mu}\}$. To find
the MPP and its probability we use an adaptation of the
Dijkstra algorithm \cite{dijkstra1959note} which takes into
account the layered and directed structure of our time-ordered
flow graph. The simplest implementation of the algorithm would
involve finding maxima by searching over the full network,
which can be a computationally expensive task. This is greatly
facilitated by using the concepts of \emph{accessibility} and
accessibility matrices \cite{lentz2013unfolding}. Thus, for
given $I$ and $J$, our implementation of the algorithm consists
of two main parts. In the first part one builds the tables
$\mathbf{U}^{(l)}_{IJ}$ of nodes accessible from $I$ and $J$ at
time step $l$, i.e. the set of nodes which can be crossed at
$t=t_l$ coming from $I$ and proceeding towards $J$ (see Fig.
\ref{fig:algorithm}). Technically, this is done by including in
$\mathbf{U}^{(l)}_{IJ}$ the nodes $k_l$ for which the two
following conditions are satisfied:
\begin{equation}
\biggl[\prod_{i=1}^{l}\mathbf{A}^{(i)}\biggl]_{Ik_l}\neq 0\qquad
\textrm{and} \qquad \biggl[\prod_{i=l+1}^{M}\mathbf{A}^{(i)}\biggl]_{k_lJ}\neq 0.
\label{accessibility}
\end{equation}

In the second part of the algorithm one recognizes that the
structure of expression (\ref{probability}) allows to maximize
it by recursively maximizing over $k_1, k_2, ..., k_{M-1}$.
This is done by finding, for each accessible node
$k_l\in\mathbf{U}^{(l)}_{IJ}$ (and only for them, without the
need of scanning the remaining nodes in the full network), the
highest probability $P^l_{I k_{l}}$ of the path connecting $I$
and $k_l$ and the actual path associated. For $l=1$, i.e. for
the first time step, trivially we have $P^1_{I k_{1}} =
\mathbf{T^{(1)}}_{I k_1}$. For $l=2,3,...,M-1$ we apply
recursively the formula
\begin{equation}
P^{l+1}_{I k_{l+1}} = \max_{k_l}\bigl(P^{l}_{I k_{l}}\mathbf{T^{(l+1)}}_{k_l k_{l+1}}\bigl).
\label{recursive}
\end{equation}
until the final point $k_M = J$ is reached, and the maximum
probability, together with the associated path, are obtained
(See Fig. \ref{fig:algorithm}). The same procedure can then be
applied to any other pair of nodes $(I',J')$.

%fig 1

Raising the number $M$ of steps we observe a fast increase in
the number of paths connecting two given nodes. It is thus
crucial to understand how much the MPP is representative of the
large set of possible paths joining two nodes. To assess in a
quantitative way this issue we introduce the following quantity
\begin{equation}\label{lambdaMPP}
\lambda_{IJ}^M = \frac{P_{IJ}^M}{\sum_{\mu}(p_{IJ}^M)_{\mu}},
\end{equation}
which determines the fraction of probability carried by the MPP
with respect to the sum of probabilities of all paths
connecting nodes $I$ and $J$. Note that the denominator can be
simply computed as the matrix-product entry
$\left(\prod_{l=1}^M \mathbf{T}^{(l)}\right)_{IJ}$.

\section{Sets of Highly probable paths}
\label{sec:hpps}

For large values of $M$, the MPP progressively loses dominance
and, on average, does not carry a significantly high fraction
of probability. However the dynamics, characterized by a high
number of paths connecting initial and final points, can be
still described by a few of them, which together have a
non-negligible probability. To see this we can relax the
definition of MPP and define a family of subsets of highly
probable paths (HPP) holding most of the probability. In our
formulation each subset ${\mathcal K}_{IJ}^M(r,\epsilon)$ is
characterized by a \emph{rank} $0 \leq r \leq M-1$ and a
threshold parameter $0\leq\epsilon\leq1$. Ideally the sets
would contain all the paths whose probability is larger than
$\epsilon P^{M}_{IJ}$. But since exhaustive searching of all
such paths becomes computationally prohibitive except for very
small $M$, the second parameter $r$ is introduced to determine
the number of constraints imposed in the search for these
relevant paths. Given the initial ($I$) and final ($J$) points
we fix $r$ nodes at intermediate times and look for paths
between $I$ and $J$ made of segments which are MPP's connecting
these intermediate nodes, by using the algorithm above.
Different locations and times for these $r$ intermediate nodes
are scanned and paths with probability larger than $\epsilon
P^{M}_{IJ}$ are retained and incorporated into the set
${\mathcal K}_{IJ}^M(r,\epsilon)$. For $\epsilon\rightarrow 1$,
independently on the rank (or for $r = 0$) only the MPP is
retained. ${\mathcal K}_{IJ}^M(r=M-1,\epsilon)$ contains all
the paths with probability larger than $\epsilon P^{M}_{IJ}$.
However, evaluation of these sets of HPPs can be
computationally costly for high values of $r$, since the
algorithm scales exponentially with $r$. Nevertheless
interesting results can be obtained considering already
low-order HPPs, i.e. $r = 1$ and $r = 2$.

Once one of the subsets is computed we can establish its
significance by defining an extension of expression
Eq.~(\ref{lambdaMPP}):
\begin{equation}\label{lambdaHPP}
\lambda_{IJ}^M(r,\epsilon)=\frac{\sum_\nu (p^{M}_{IJ})_\nu}{\sum_{\mu}(p^{M}_{IJ})_{\mu}},
\end{equation}
where the sum in the numerator is over all the paths in the
subset ${\mathcal K}_{IJ}^M(r,\epsilon)$ and the one in the
denominator is over all paths connecting $I$ to $J$.

Another important aspect of the sets of HPPs is to establish
how close, spatially, are they with respect to the
corresponding MPP. This is obtained with an average distance
function. Given two generic paths between initial and final
points $I$ and $J$, $\mu_1=\{I,k_1...,J\}$ and
$\mu_2=\{I,l_1...,J\}$ we define their average distance as
\begin{equation}\label{Distancefunction}
\mathbf{d}(\mu_1,\mu_2)=\frac{1}{M-1}\sum_{i=1}^{M-1}d(k_i,l_i),
\end{equation}
where $d(k_i,l_i)$ is a metric determining the distance between
two given nodes of the network. For a geophysical transport
network the geographical distance (on the sphere) between the
centers of the nodes is the most natural choice. For a given
pair of nodes $(I,J)$ the average distance between the subset
${\mathcal K}_{IJ}^M(r,\epsilon)$ and the MPP connecting them
in $M$ time steps is defined as
\begin{equation}\label{Averagedistance}
\mathcal{D}^{M}_{IJ}=\frac{1}{N^{M}_{IJ}}\sum_{\mu}\mathbf{d}(\mu,\eta_{IJ}^{M}),
\end{equation}
where $N^{M}_{IJ}$ is the number of paths $\mu$ in the subset
${\mathcal K}_{IJ}^M(r,\epsilon)$, and the sum is extended over
all paths in the subset (remember that $\eta_{IJ}^{M}$ denotes
the MPP). This quantity provides an estimation
 of how much paths in the subset
deviate spatially from the correspondent MPP. A large deviation
means that the probability to reach $J$ from $I$ is spatially
spread in a large region and indicates furthermore the
importance of considering the HPP subset instead of only the
MPP. Small values of $\mathcal{D}^{M}_{IJ}$ imply HPP sets with
the shape of coherent narrow tubes around the MPP, so that the
MPP already characterizes the spatial pathways, even if its
probability is not large.

In the next Sections we apply the above formalism to the
atmospheric flow occurring over Eastern Europe in Summer 2010.
Computations of optimal paths and their sets in an analytic
double-gyre system, a much simpler flow in which path
properties could be more easily appreciated, are contained in
the Appendix.

\section{A network of atmospheric flow over eastern Europe in Summer 2010}
\label{sec:atm}

In this section we describe the physical characteristics of the
atmospheric event, the data used and the model we employ to
obtain the air particle trajectories.

\subsection{Event description}
\label{subsec:event}

\begin{figure}
\centering
\includegraphics[width=\columnwidth]{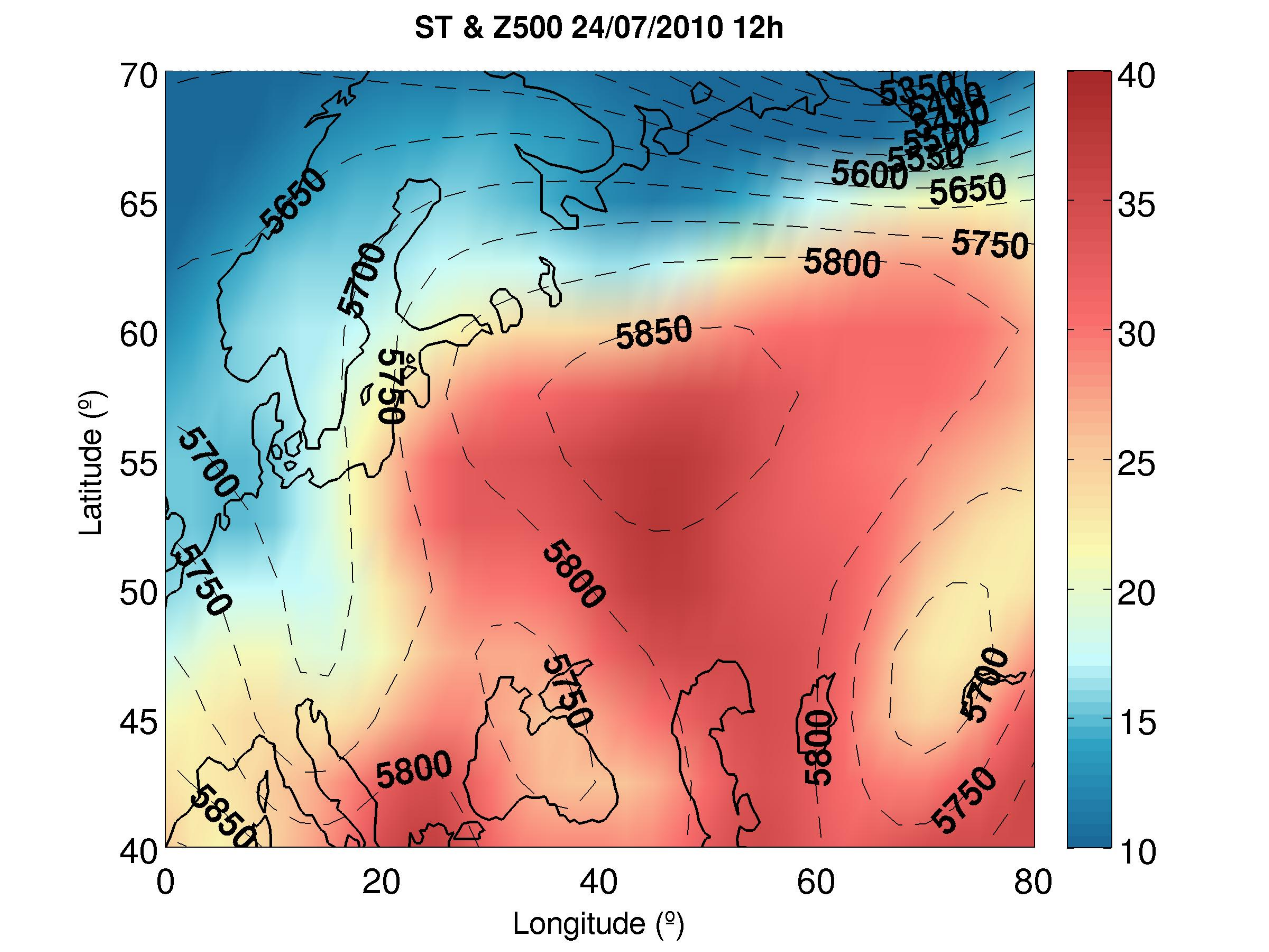}
\caption{Geopotential height at 500hPa (contours, in m)
and temperature (color code, in degrees C) over the region of interest, on
July 24th, 12:00 UTC.}
\label{fig:Eulerian}
\end{figure}

Eastern Europe and Western Russia experienced a strong,
unpredicted, heat wave during the summer of 2010. Extreme
temperatures resulted in over 50000 deaths and inflicting large
economic losses to Russia. The heat wave was due to a strong
atmospheric blocking that persisted over the Euro-Russian
region from late June to early August
\cite{matsueda2011predictability}. During July the daily
temperatures were near or above record levels and the event
covered Western Russia, Belarus, Ukraine, and the Baltic
nations. Physically, the origins of this heat wave were in a
atmospheric block episode that produced anomalously stable
anticyclonic conditions, redirecting the trajectories of
migrating cyclones. Atmospheric blocks can remain in place  for
several days (sometimes even weeks) and are of large scale
(typically larger than $2000$ km). In particular, the Russian
block of summer 2010 was morphologically of the type known as
{\it Omega block} that consists in a combination of
low-high-low pressure fields with geopotential lines resembling
the Greek letter $\Omega$ (see Fig.\ref{fig:Eulerian}). Omega
blocks bring warmer and drier conditions to the areas that they
impact and colder, wetter conditions in the upstream and
downstream \cite{black200factors}. We study the concrete period
extended from the July the 20th to July 30th.

% figure 2

\subsection{Data}
\label{subsec:data}

Atmospheric data were provided by the National Centers for
Environmental Prediction (NCEP) Climate Forecast System
Reanalysis (CFSR) through the Global Forecast System (GFS)
\cite{saha2010ncep}. This reanalysis was initially completed
over the $31$ year period from $1979$ to $2009$ and extended to
March $2011$. Data can be obtained with a temporal resolution
of $1$ hour and a spatial horizontal resolution of
$0.5\degree\times 0.5 \degree$. The spatial coverage contains a
range of longitudes of $0 \degree E$ to $359.5 \degree E$ and
latitudes of $90 \degree S$ to $90 \degree N$.

The variables needed as input to the Lagrangian dispersion
model described in the next section include dew point
temperature, geopotential height, land cover, planetary
boundary layer height, pressure and pressure reduced to mean
sea level, relative humidity, temperature, zonal and meridional
component of the wind, vertical velocity and water equivalent
to accumulated snow depth. All these fields are provided by
CFSR data on $26$ pressure levels.

\subsection{Lagragian Particle Dispersion Model FLEXPART}
\label{subsec:flexpart}

As mentioned, the idea is to obtain the effective velocity
field felt by any fluid particle. Then the Lagrangian
dispersion model (see next subsection) will integrate it to
provide as output the three-dimensional positions of the
particle at every time step.

The numerical model used to integrate particle velocities and
obtain trajectories is the Lagrangian particle dispersion model
FLEXPART version $8.2$
\cite{stohl2005technical,stohl2011lagrangian}. FLEXPART
simulates the long-range and mesoscale transport, diffusion,
dry and wet deposition, and radioactive decay of tracers
released from point, line, area or volume sources. It most
commonly uses meteorological input fields from the numerical
weather prediction model of the European Centre for
Medium-Range Weather Forecasts (ECMWF) as well as the Global
Forescast System (GFS) from NCEP (the one used in our study).
Trajectories are produced by integrating the equation (the
input velocity data are interpolated on the present particle
position):
\begin{equation}
\frac{d{\bf X}}{dt}= {\bf v} ({\bf X }(t)),
\label{pteq}
\end{equation}
with $t$ being time, $\bf X$ the vector position of the air
particle, and ${\bf v} = {\bf \bar v}+{\bf v^t} +{\bf v^m}$ is
the wind vector. FLEXPART takes the grid scale wind ${\bf
\bar v}$ from the CFSR, but complements it with stochastic components
${\bf v^t}$ and ${\bf v^m}$ to better simulate the unresolved
turbulent processes occurring at small scales. The turbulent
wind fluctuations ${\bf v^t}$ are parametrized by assuming a
Markov process via a Langevin equation, and the mesoscale wind
fluctuations ${\bf v^m}$ are implemented also via an
independent Langevin equation by assuming that the variance of
the wind at the grid scale provides information on the subgrid
variance. Variables entering the parametrizations are obtained
from the meteorological CFSR fields. For additional details we
refer to Stohl et al.
\cite{stohl2005technical,stohl2011lagrangian}.

\subsection{Network construction}
\label{subsec:net}

\begin{figure}
\centering
\includegraphics[width=\columnwidth, clip=true]{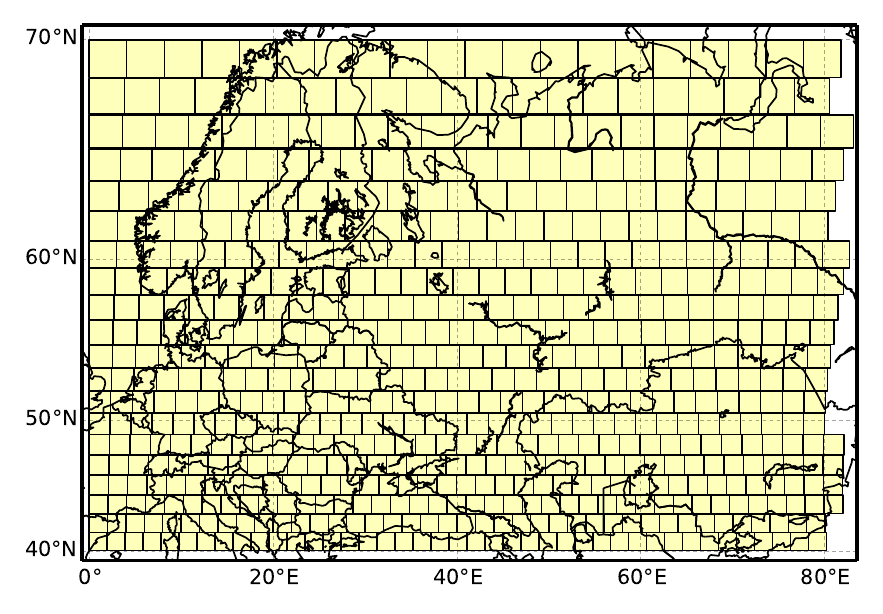}
\caption{The geographical domain considered and the
discretization grid defining the nodes of our flow network.}
\label{fig:grid}
\end{figure}

We focus our analysis on the domain in between $0^\circ$E -
$80^\circ$E and $40^\circ$N - $70^\circ$N. In order to define
the nodes of the network we discretize this region in $626$
equal-area boxes using a sinusoidal projection. The latitudinal
extension of each node-box is $1.5^\circ$, the longitudinal one
varies depending on the latitude (see Fig.~\ref{fig:grid}). The
area of each box is 27722 $km^2$, so that the typical
horizontal size is of the order of 166.5 $km$. This is a
moderate coarse-graining of the resolution ($0.5^\circ\times
0.5 ^\circ$) of the NCEP data used for particle integration. We
take $\tau=12$ hours as time discretization, which is enough to
follow the dynamics of the blocking event. It has been shown in
an oceanic flow network\cite{sergiacomi2015most} that the value
of $\tau$ has a minor influence on optimal paths, being more
important the total time-interval considered $M\tau$. We
uniformly fill each node with $800$ ideal fluid particles
releasing them at $5000\ m$ of height, a representative level
in the middle troposphere. FLEXPART trajectories are fully
threedimensional, but by initializing at each time-step
particles in a single layer we are effectively neglecting the
vertical dispersion (which is of the order of 800 m in the
$\tau=12~h$ time step) and focussing on the pathways of large
scale horizontal transport. Fully three-dimensional flow
networks will be the subject of future work.

% figure 3

\section{Results}
\label{sec:results}

\subsection{Optimal paths}

\begin{figure}
\centering
\includegraphics[width=\columnwidth,clip=true]{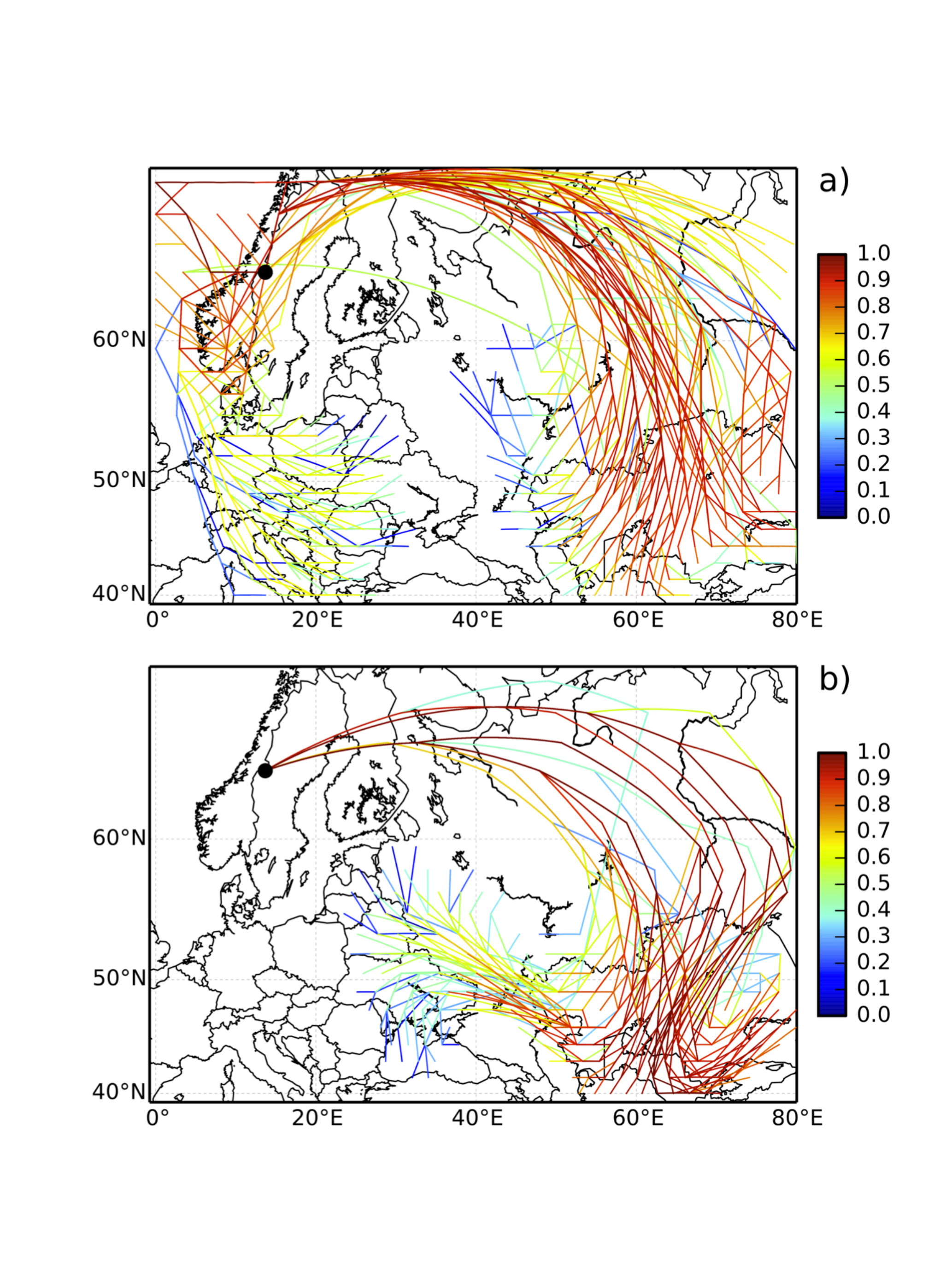}
\caption{Paths of M = 9 steps of $\tau=12$ hours in our flow
network with starting date July 25th 2010 (panel a)) and July
20th 2010 (panel b)), represented as straight segments (in
fact, maximal arcs on the Earth sphere) joining
the path nodes. MPPs originating from a single node (black
circle) and ending in all accessible nodes. Color gives the
$P_{IJ}^{M}$ value of the paths in a normalized log-scale
between the minimum value (deep blue) and the maximum (dark
red). Panel a): probabilities ranging from $10^{-3}$ to
$10^{-14}$. Panel b): probabilities ranging from $10^{-3}$ to
$10^{-15}$.}
\label{fig:pathsmain}
\end{figure}

Equipped with the tools developed above we can now compute
pathways of transport during the atmospheric event described in
Sect. \ref{sec:atm}. Figure \ref{fig:pathsmain}a shows all the
optimal paths leaving a node in the Scandinavian Peninsula at
July 25 and arriving to all nodes which are reached in $M=9$
steps (i.e. 4.5 days). The graphical representation joins with
maximal arcs the center of the grid boxes identified as
pertaining to the MPP. The actual particle trajectories between
two consecutive boxes are not necessarily such arcs. The paths
are colored according to their probability value $P_{IJ}^M$.
The MPPs with highest probability (reddish colors) follow a
dominant anticyclonic (i.e. clockwise) route bordering the high
pressure region (see Fig. \ref{fig:Eulerian}, but note that
this is at a particular time, whereas the trajectory plots span
a range of dates of more than four days) without penetrating
it. There is also a branch of MPPs with much smaller
probabilities (yellow and bluish colors) that are entrained
southward by a cyclonic circulation.

%figure 4

\begin{figure}
\centering
\includegraphics[width=\columnwidth,clip=true]{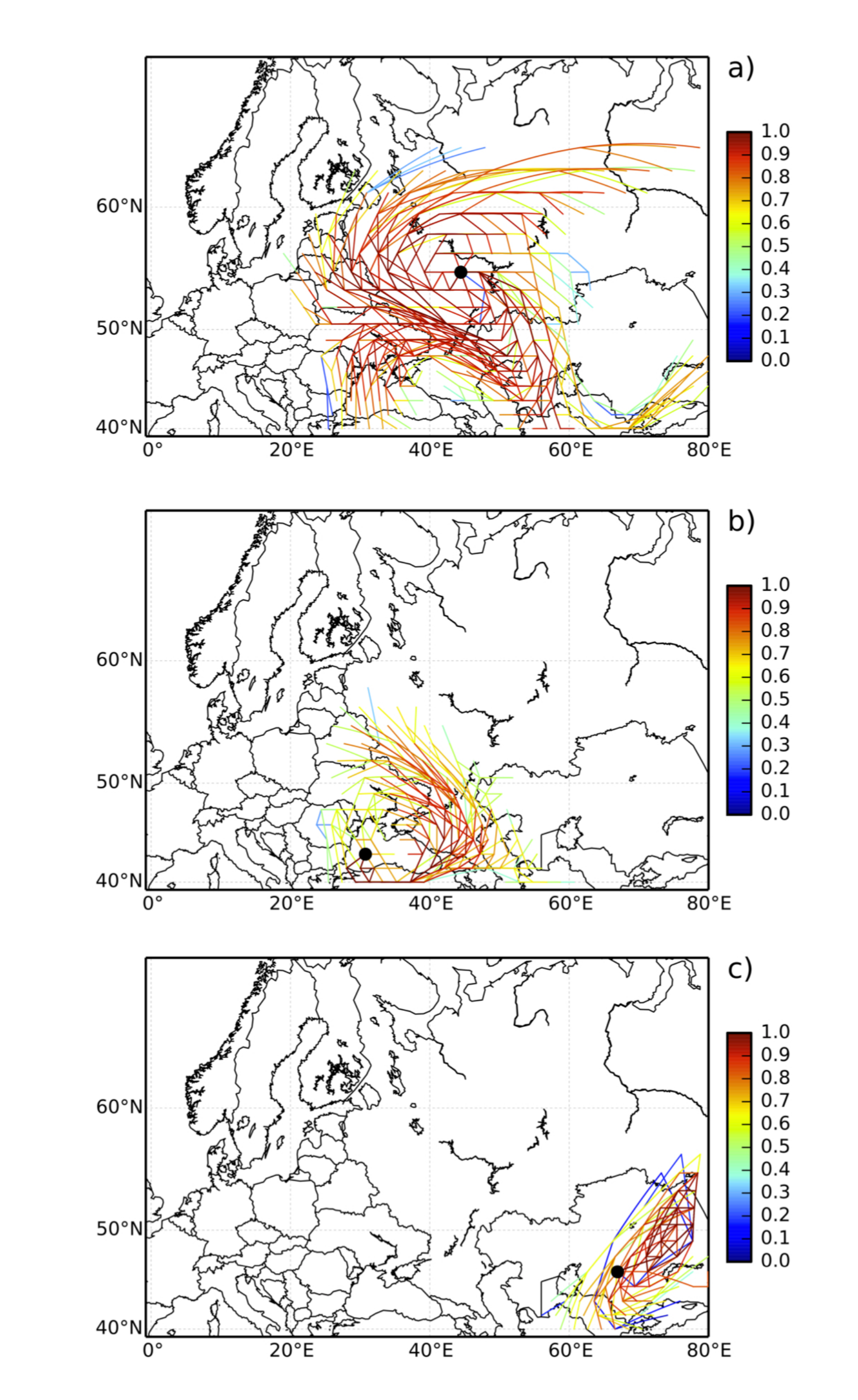}
\caption{Optimal paths of 9 steps of $\tau=12$ hours with
starting date July 20th 2010, entrained in the high- and in the two low-pressure
areas of the blocking. Same coloring scheme as in Fig. \ref{fig:pathsmain}.
Panel a): probabilities
ranging from $10^{-3}$ to $10^{-16}$. Panel b): probabilities
ranging from $10^{-2}$ to $10^{-16}$. Panel c): probabilities
ranging from $10^{-3}$ to $10^{-13}$.}
\label{fig:paths3gyres}
\end{figure}

Despite the persistent character of the Eulerian block
configuration, sets of Lagrangian trajectories become highly
variable in time. See for example the set of MPPs starting from
the same initial location but five days earlier (Fig.
\ref{fig:pathsmain}b). The southward cyclonic branch is now
absent, all MPPs following initially the anticyclonic gyre.
Remarkably, the set of trajectories bifurcates into two
branches when approaching what seems to be a strong hyperbolic
structure close to 40$^\circ$N 60$^\circ$E. A hint of the
presence of second hyperbolic structure is visible at the end
of the westward branch, close to 50$^\circ$N 30$^\circ$E.
Figure \ref{fig:paths3gyres} displays additional MPPs starting
also at July 20th, but initialized inside the main anticyclonic
region of the blocking, and in two low-pressure regions
flanking it. Fig. \ref{fig:paths3gyres}a clearly shows the main
anticyclonic circulation, highlighting also the escape routes
from the high-pressure zone, associated with the hyperbolic
regions described above. The other two panels show the cyclonic
circulations at each side of the high, in a characteristic
Omega-blocking configuration. It is remarkable the compactness
of the trajectories inside the eastern low-pressure area, which
form a very localized and coherent set with practically no
escape in the 4.5 days time-interval displayed.

%figure 5
%\begin{widetext}
%\end{widetext}

We stress that the plots in Figs. \ref{fig:pathsmain} and
\ref{fig:paths3gyres} are different from \emph{spaghetti plots}
for which many available trajectories are plotted from
different or related initial conditions. For our set of
particles this will give 800 trajectories emanating from each
box. Here we are plotting just one path, the MPP, for each
initial and final box pair, which strongly limits the number of
paths from each box but, as we will see more thoroughly, it is
still representative of the trajectories of many released
particles.

\subsection{Relevance of the MPPs}

%figure 6
\begin{figure}
\centering
\includegraphics[width=\columnwidth]{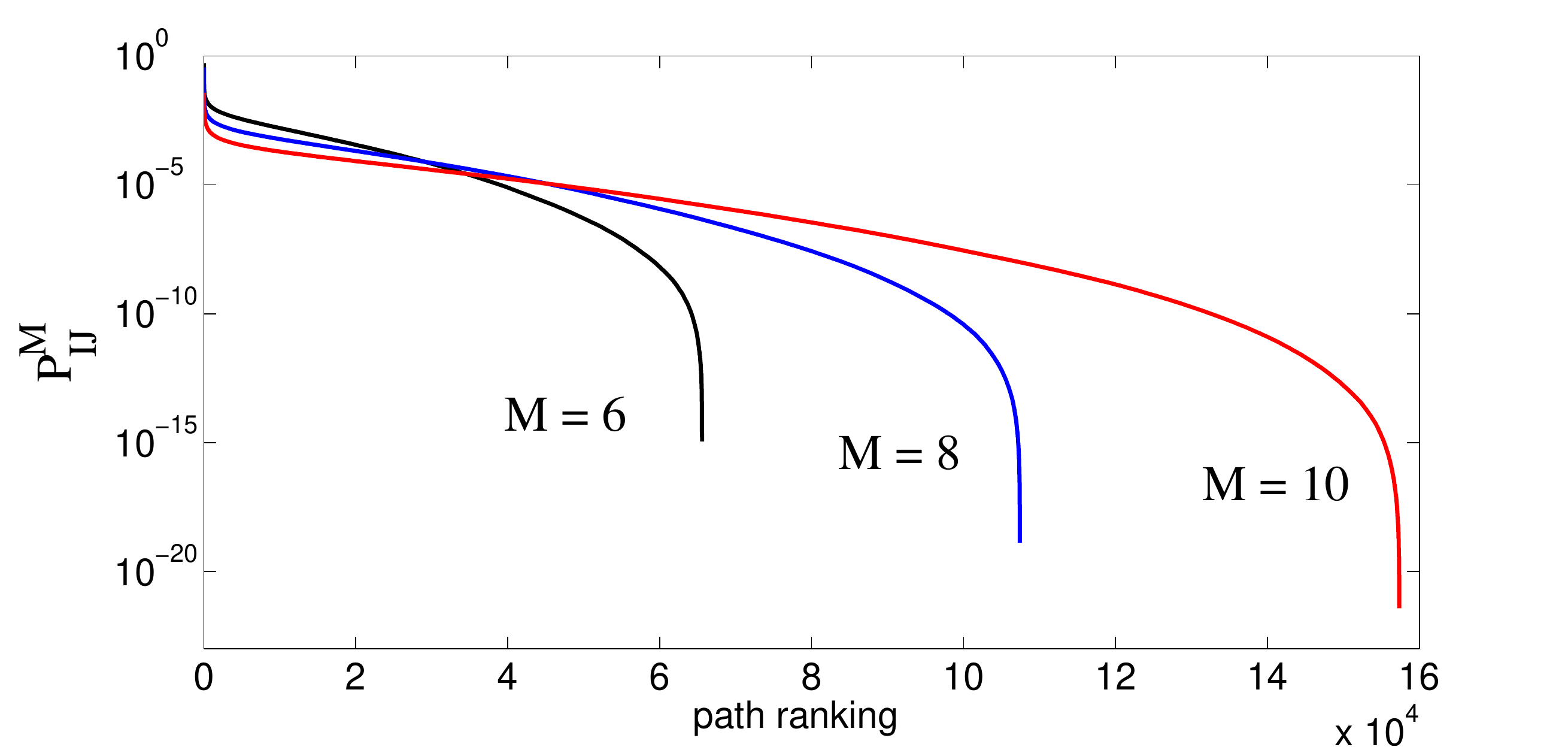}
\caption{Ranking plot in which the $P_{IJ}^M$ values of all MPPs obtained for
$M=6$,8, and 10 starting on July 25th in the whole area are plotted in decreasing order.
The range of probability
values of the MPPs can be read from the vertical axis (from a few percent to
$10^{-15}$ for $M=6$ or to less than $10^{-20}$ for $M=10$). The total number of optimal paths
can also be read-off from the horizontal axis.}
\label{fig:PIJ}
\end{figure}

\begin{figure}
\centering
\includegraphics[width=.8\columnwidth]{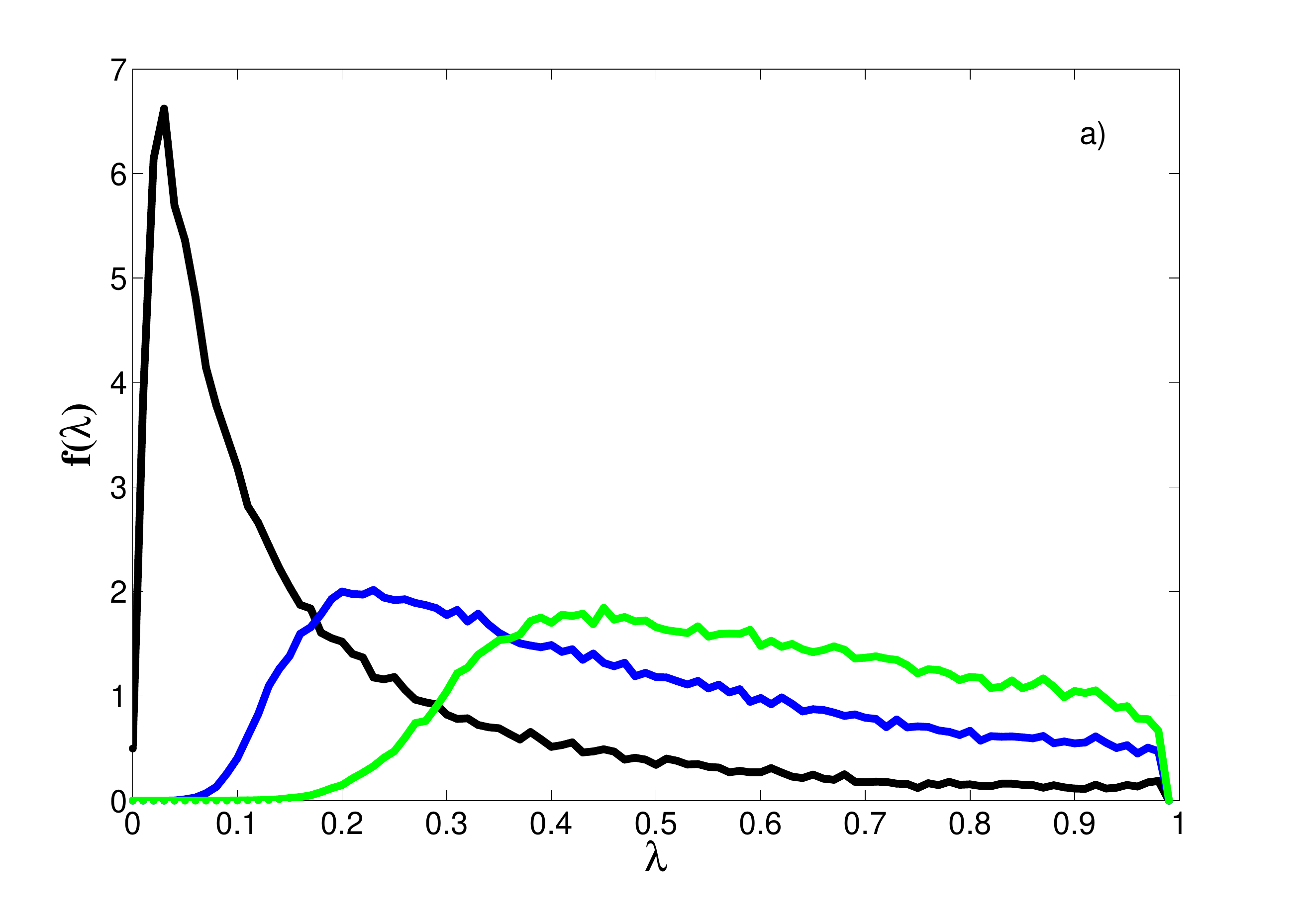}
\includegraphics[width=.8\columnwidth]{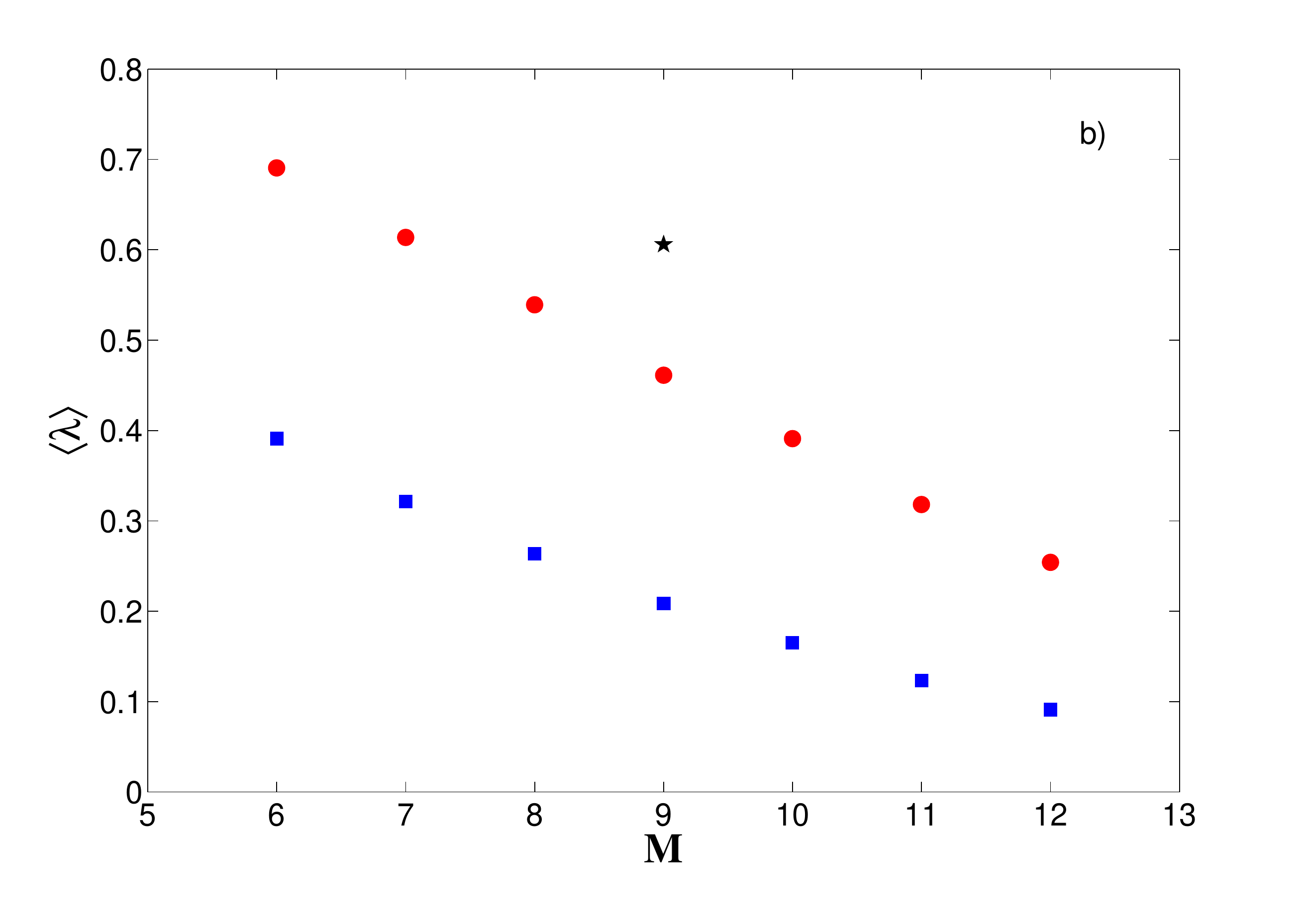}
\caption{a) Normalized probability density $f(\lambda)$
of the merit figure $\lambda_{IJ}^M(r,\epsilon)$ of paths started on July 20th 2010 for
M = 9 and $\epsilon = 0.1$,
with $r = 0$ (only the MPPs, black curve), $r = 1$ (blue) and $r = 2$ (green).
b) Mean value of the $\lambda_{IJ}^M(r,\epsilon)$ distributions (paths' starting date July 25th) as
a function of the number of time steps $M$ for $r = 0$ (only MPPs, blue squares), $r = 1$
(red circles) and $r = 2$ (single black star). }
\label{fig:lambda}
\end{figure}

The range of colors in Figs. \ref{fig:pathsmain} and
\ref{fig:paths3gyres} indicates that, given an initial box, not
all MPPs leading to different locations are equally probable.
This is quantified by the probability $P_{IJ}^M$ which gives a
weight to each MPP. Indead $P_{IJ}^M$ takes a very large range
of values.  Figure \ref{fig:PIJ} shows a ranking plot in which
the values of all MPPs of a given $M$ and started at a
particular date are plotted in decreasing order. We see a huge
spread on the values of $P_{IJ}^M$. Very low probability values
arise because of the exponential explosion of the number of
paths between two nodes with increasing $M$. Given these low
values of $P_{IJ}^M$ except for the smallest values of $M$, one
should ask how representative are the MPPs for the full set of
paths. Figure \ref{fig:lambda}a shows distributions of the
parameter $\lambda_{IJ}^M(r,\epsilon)$ giving the relative
importance of the different types of paths. We see that
$\lambda$-values are small when considering only the MPPs
($r=0$), but the distributions shift towards higher values for
paths sets of increasing $r$. Figure \ref{fig:lambda}b gives
mean values of the $\lambda$ distributions. They decrease with
$M$, reflecting the lack of representativeness of the smallest
sets of paths for large $M$. However, already for $r=1$ the set
of HPPs has a mean value higher than 0.5 for a relevant range
of time steps.

%figure 7

\begin{figure*}
\centering
\includegraphics[width=\textwidth,clip=true]{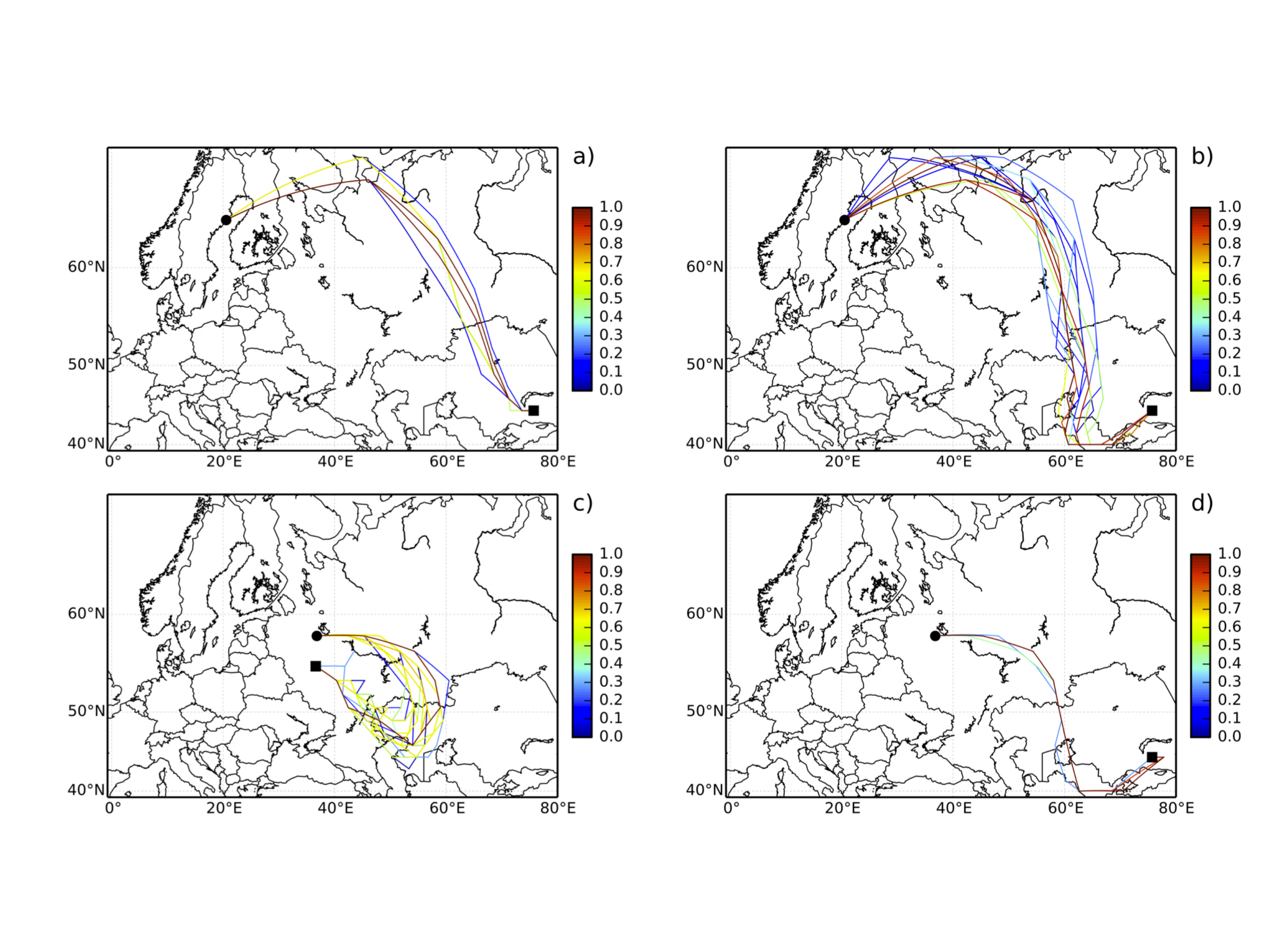}
\caption{All paths in $\mathcal{K}_{IJ}^{M}(r=1,\epsilon=0.1)$
for different $I,J$, initial point $I$ marked by a circle and
final point $J$ marked by a square. The color bar gives in
logarithmic scale values ranging from the maximum one
$P_{IJ}^M$ (dark red), corresponding to the MPP, to the minimum
of $0.1 P_{IJ}^M$. Panel a): $M=8$ steps, with starting date
July 25th 2011; $P_{IJ}^M=7.8\times 10^{-5}$.  Panel b): $M=12$
steps, with starting date July 25th 2011; $P_{IJ}^M=2.7\times
10^{-5}$. Panel c): $M=11$ steps, with starting date July 20th
2011; $P_{IJ}^M=7.4\times 10^{-7}$. Panel d): $M=11$ steps,
with starting date July 20th 2011; $P_{IJ}^M=1.5\times
10^{-7}$. }
\label{fig:tubes}
\end{figure*}

Thus, for the values of $M$ and $\epsilon$ discussed here, the
set of HPPs with $r=1$ seems to be rich enough to represent the
transport pathways. But how different is the geometry of the
different paths in this HPP set? And how different is it from
the MPPs? We plot in Fig. \ref{fig:tubes} examples of all HPPs
with $r=1$ and $\epsilon=0$ for particular $(I,J)$ values and
dates. In all the cases the sets remain coherent and narrow
tubes of trajectories defining roughly the same pathway as the
MPP.

%figure 8

\begin{figure}
\centering
\includegraphics[width=\columnwidth, clip=true]{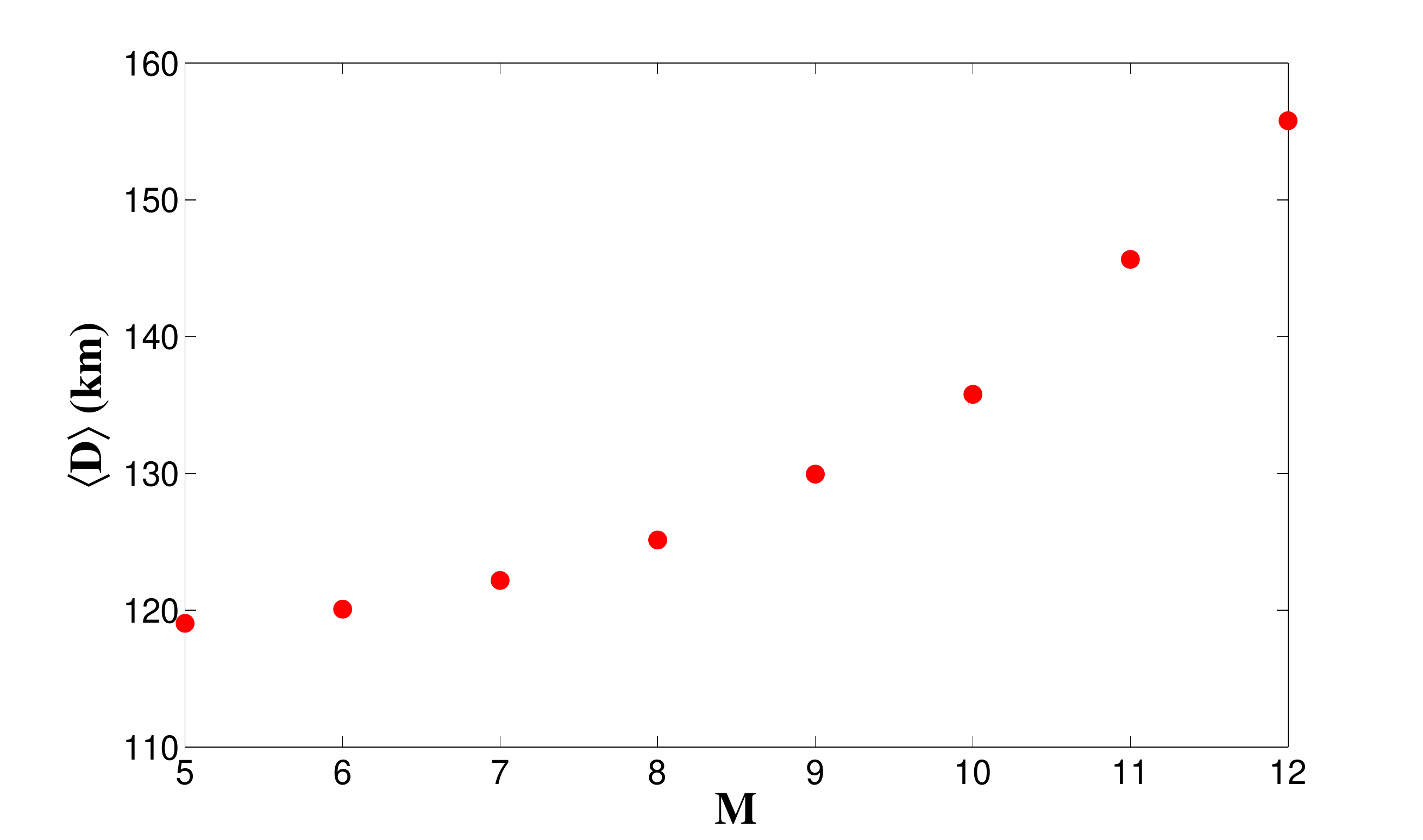}
\caption{Plot of the mean distance
$\mathcal{D}^{M}_{IJ}$ (Eq. (\ref{Averagedistance})) as a function of $M$
for $r = 1$ and $\epsilon=0.1$. The quantity is further averaged over all the HPPs
starting on July 25th. Units are kilometers.}
\label{fig:distance}
\end{figure}

A quantification of the \emph{width} of the tubes can be done
with the distance measure $\mathcal{D}^{M}_{IJ}$ in Eq.
(\ref{Averagedistance}). An average of it over pairs of
locations is shown in Fig. \ref{fig:distance}. Although the
tube width increases with $M$, it remains always below the
typical linear box size of approximately 166.5 km (see Sect.
\ref{subsec:net}) indicating that the tubes remain narrow. Thus
we conclude that, despite the decreasing probability of the
MPPs for increasing $M$, they remain good indicators of the
dominant pathways in the transport network.

%figure 9

\begin{figure}
\centering
\includegraphics[width=\columnwidth, clip=true]{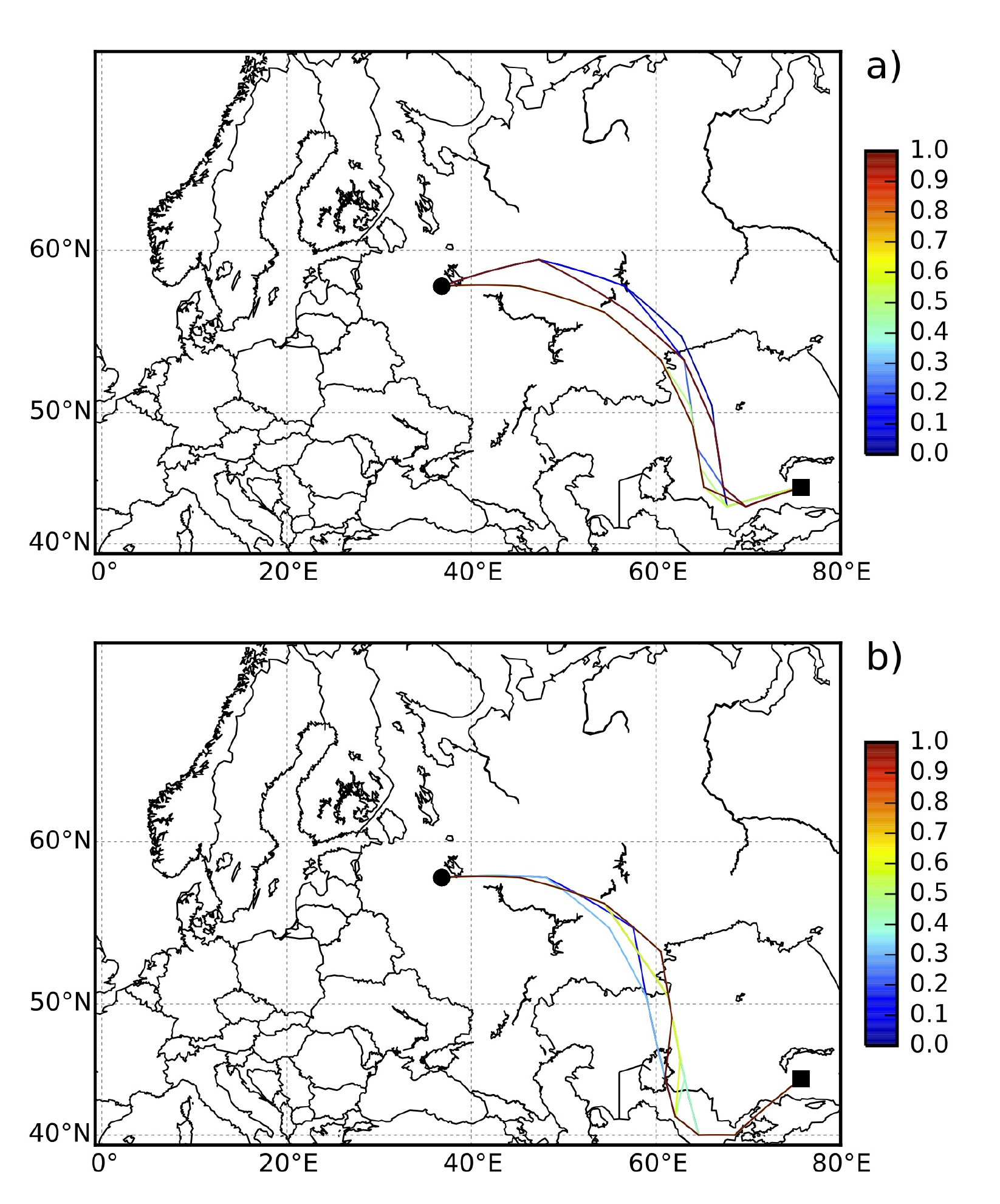}
\caption{All paths in $\mathcal{K}_{IJ}^{M}(r=1,\epsilon=0.1)$
for the same $I,J$ and starting date as in Fig. \ref{fig:tubes}d. Same
coloring scheme as in Fig.\ref{fig:tubes}. Panel
a) $M=7$ steps; $P_{IJ}^M=2.2\times 10^{-5}$. b) $M=9$ steps;
$P_{IJ}^M=2.3\times 10^{-4}$.}
\label{fig:timedependence}
\end{figure}

As a final description of properties of the dominant transport
paths, we present in Figure \ref{fig:timedependence} (compare
with Fig. \ref{fig:tubes}d) an example on how the MPP and the
HPPs between a fixed pair of nodes change when considering
different values of $M$, defining the temporal interval.
Typically, the probability of the MPP shows a maximum at some
intermediate value of $M$ in between shorter values of $M$ for
which very few particles connect the two nodes, and larger
values of $M$ for which the increasing number of factors
smaller than one in the product (\ref{probability}) defining
$(p^M_{IJ})_{\mu}$ makes this quantity to decrease again until
vanishing. For the example shown in Figs. \ref{fig:tubes}d and
\ref{fig:timedependence}, the value of $M$ giving the maximum
$P_{IJ}^M$ is around $M\approx 9$, i.e. $M\tau=4.5$ days. Note
that the HPP trajectories change length but keep a similar
shape in the range of $M$ considered, indicating that in this
time interval the blocking atmospheric structures evolve
slowly.

% figure 10

%\textbf{NO:} Should we discuss \textbf{betweenness}? The first
%impression from the plots is that high betweenness is locating
%hyperbolic structures. Thus they should be related to the FSLE,
%which in Irene's thesis are only backwards.
%
%\begin{figure*}
%\centering
%%\includegraphics[width=\columnwidth, clip=true]{fig1.eps}
%\includegraphics[width=0.4\textwidth]{paths_sum_Subset_2010072001_11h_09step_r=0_001eps_FromAll_ToAll.png}
%\includegraphics[width=0.4\textwidth]{paths_sum_Subset_2010072001_11h_12step_r=0_001eps_FromAll_ToAll.png}
%\includegraphics[width=0.4\textwidth]{paths_sum_Subset_2010072001_11h_18step_r=0_001eps_FromAll_ToAll.png}
%\includegraphics[width=0.4\textwidth]{paths_sum_Subset_2010072001_11h_24step_r=0_001eps_FromAll_ToAll.png}
%\caption{Betweenness for 9,12,18 and 24 steps}
%\label{fig:betweennness}
%\end{figure*}
%
%\textbf{NO:} Can we see some plot of \textbf{entropy} (forw and
%back) to see if we see something?

%%%%%%%%%%%%%%%%%%%%%%%%%%%%%%%%%%%%%%%%%%%%%%%%%%%%%%%%%%%%%

\section{Conclusions}
\label{sec:conclusions}

We have introduced MPPs and sets of HPPs as tools to visualize
and analyze dominant pathways in geophysical flows. We have
computed them for an atmospheric blocking event involving
eastern Europe and Western Russia. The computed optimal paths
give a Lagrangian view of the Omega-block configuration, with a
central anti- cyclonic circulation flanked by two cyclonic
ones. Moreover they give additional insight on it, such as the
variability of the dominant pathways, and the identification of
escaping and trapping regions. The statistical significance of
single MPPs decreases with the time interval considered, but we
find always that the MPPs remain representative of the spatial
geometry of the pathways, in the sense that the sets of HPPs
are coherent narrow tubes providing transport paths always
close to the optimal path. This spatial coherence of transport
between pairs of locations was already noticed in an ocean flow
\cite{sergiacomi2015most} and it is also present in the model
flow discussed in the Appendix. Then, it seems to be a general
characteristic of flow networks.

% If you have acknowledgments, this puts in the proper section head.
\begin{acknowledgments}
We acknowledge financial support from FEDER and MINECO (Spain)
through the ESCOLA (CGrant no. TM2012-39025-C02-01) and
INTENSE@COSYP (Grant no. FIS2012-30634) projects, and from the
European Commission Marie-Curie ITN program (FP7-320
PEOPLE-2011-ITN) through the LINC project (Grant no. 289447).
\end{acknowledgments}

%\section*{Appendix A (if needed): Appendix title ...}
%\renewcommand{\theequation}{A\arabic{equation}}
%\setcounter{equation}{0}  % reset counter
%\renewcommand{\thefigure}{A\arabic{figure}}
%\setcounter{figure}{0}  % reset counter
%
%An Appendix, if needed

\section*{Appendix A: Optimal paths in a simple model system}
\renewcommand{\theequation}{A\arabic{equation}}
\setcounter{equation}{0}  % reset counter
\renewcommand{\thefigure}{A\arabic{figure}}
\setcounter{figure}{0}  % reset counter

In this Appendix we display optimal paths and sets of optimal
paths for an analytic model flow, the double-gyre. See for
example \cite{shadden2005definition,farazmand2012computing} for
basic properties of this system and computations of its
Lagrangian coherent structures and Lyapunov fields. Because of
the simplicity of this flow as compared with the atmospheric
situation studied in the main text, characteristics of the
optimal paths could be appreciated more easily.

The double-gyre is a two-dimensional time-periodic flow defined
in the rectangular region of the plane $\bx=(x,y) \in
[0,2]\times[0,1]$. It is described by the streamfunction
\BE
\psi(x,y,t)= A \sin(\pi f(x,t))\sin(\pi y) \ ,
\label{dg-stream1}
\EE
with
\BA
f(x,t)&=& a(t)x^2+b(t)x \\
a(t)  &=& \gamma \sin(\omega t) \ , \\
b(t)  &=& 1-2\gamma \sin(\omega t) \ . \label{dg-stream2}
\EA
From these expressions, the velocity field is
\BA
\dot x &=& -\frac{\partial\psi}{\partial y}=-\pi A \sin(\pi f(x,t))\cos(\pi y)\\
\dot y &=& \frac{\partial\psi}{\partial x}=\pi A \cos(\pi
f(x,t)) \sin(\pi y) \frac{\partial f(x,t)}{\partial x} \ .
\label{dg-v}
\EA

For $\gamma=0$, this flow is steady. Ideal fluid particles
follow very simple trajectories: they rotate following closed
streamlines, clockwise in the left half of the rectangle, and
counterclockwise in the right one. The central streamline
$x=1$, a heteroclinic connection between the hyperbolic point
at $(1,1)$ and the one at $(1,0)$, acts as a separatrix between
the two regions. When $\gamma > 0$, more complex behavior
including chaotic trajectories arises. The periodic
perturbation breaks the separatrix, so that now some
interchange of fluid is possible between the left and the right
part of the rectangle. The geometric structures involved in
this interchange have been studied with a variety of techniques
\cite{shadden2005definition,farazmand2012computing} but the
framework of optimal paths developed in this paper seems quite
natural for this purpose.

We take the parameters $A=0.1$ and $\omega=2\pi/5$, and compute
paths in our network framework for two qualitatively different
situations, namely the steady case $\gamma=0$, and the
periodically perturbed case (of period $2\pi/\omega=5$) with
$\gamma=0.3$. We discretize the fluid domain into
$100\times50=5000$ square boxes, defining the nodes in our flow
network, and compute the adjacency matrices $\mathbf{A}^{(l)}$,
$l=1...M$, by releasing 400 particles from each of the boxes.
In all the cases shown below we compute paths of $M=6$ steps of
duration $\tau=1$, starting at $t_0=0$.

% figure A1
\begin{figure}
\centering
\includegraphics[width=\columnwidth]{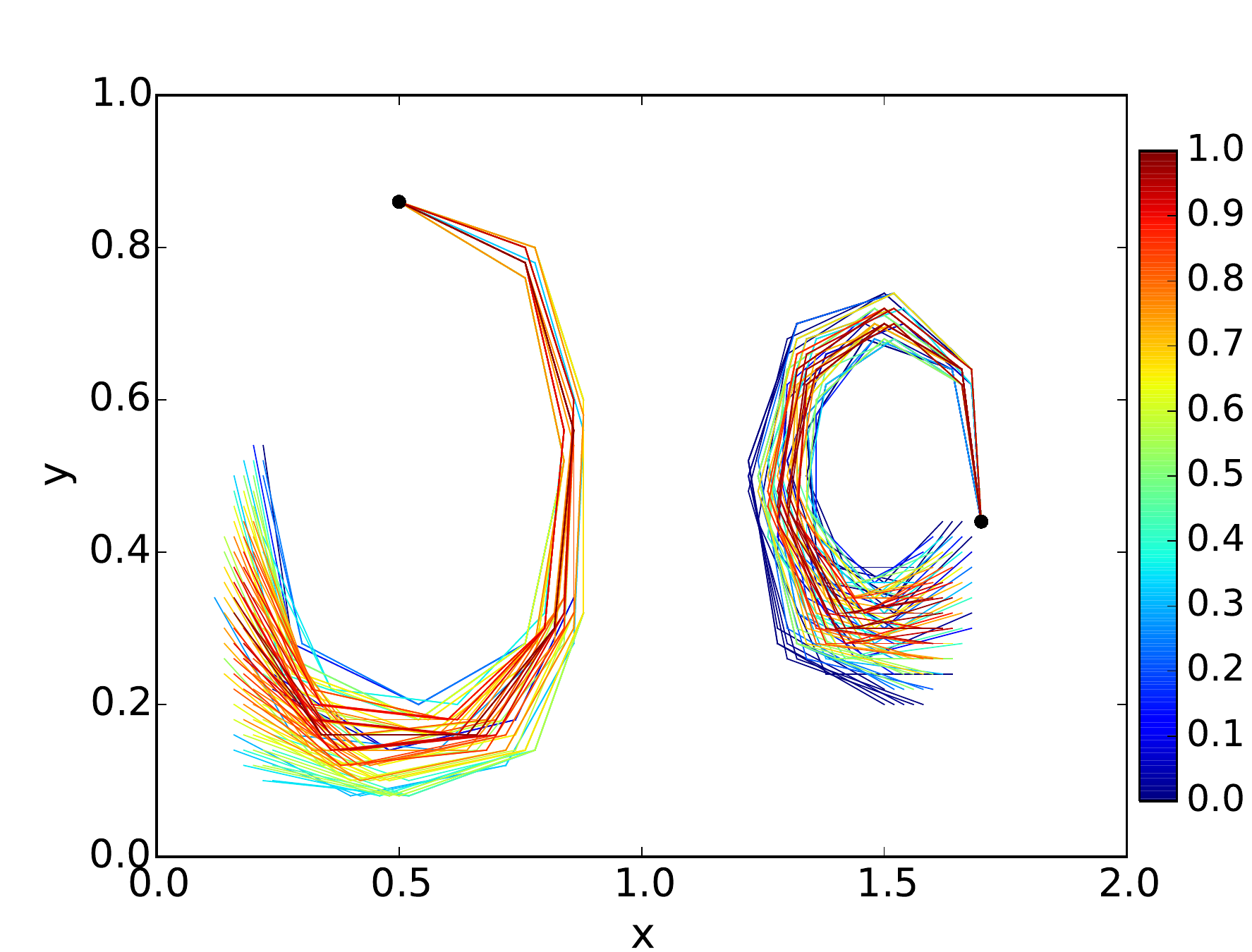}
\caption{Double-gyre 6-step paths in the steady case $\gamma=0$, from two
different starting nodes (the black circles) to all the accessible destinations.
The network nodes pertaining to each path are joined by straight line segments colored
according to the path probability. The color code is logarithmic
in the full probability range, which is
$[0.054, 7.73\times10^{-10}]$ for the left node (there is a total of 94
paths emanating from it) and $[0.0234, 10^{-7}]$ for the right one (87 paths).}
\label{fig:A1}
\end{figure}

Figure \ref{fig:A1} considers the steady flow ($\gamma=0$) and
shows all optimal paths emanating from two particular initial
nodes and reaching all nodes accessible from them after the 6
steps. We see the general clockwise and anticlockwise
circulations at each side of the separatrix. The two halves of
the domain remain isolated. Note that the paths are different
from the closed streamlines. This is so because the
discretization of the fluid domain into finite boxes, together
with the Markov assumption, introduces an stochastic component
equivalent to an effective diffusivity
\cite{froyland2013analytic} and leads to dispersion of the
particles starting from a single node. In our atmospheric
velocity flow there were in addition explicit stochastic terms
modeling turbulent diffusion and mesoscale fluctuations. Note
also that, as in the atmospheric case, a huge range of values
of $P_{IJ}^M$ is present.

% figure A2
\begin{figure}
\centering
\includegraphics[width=\columnwidth]{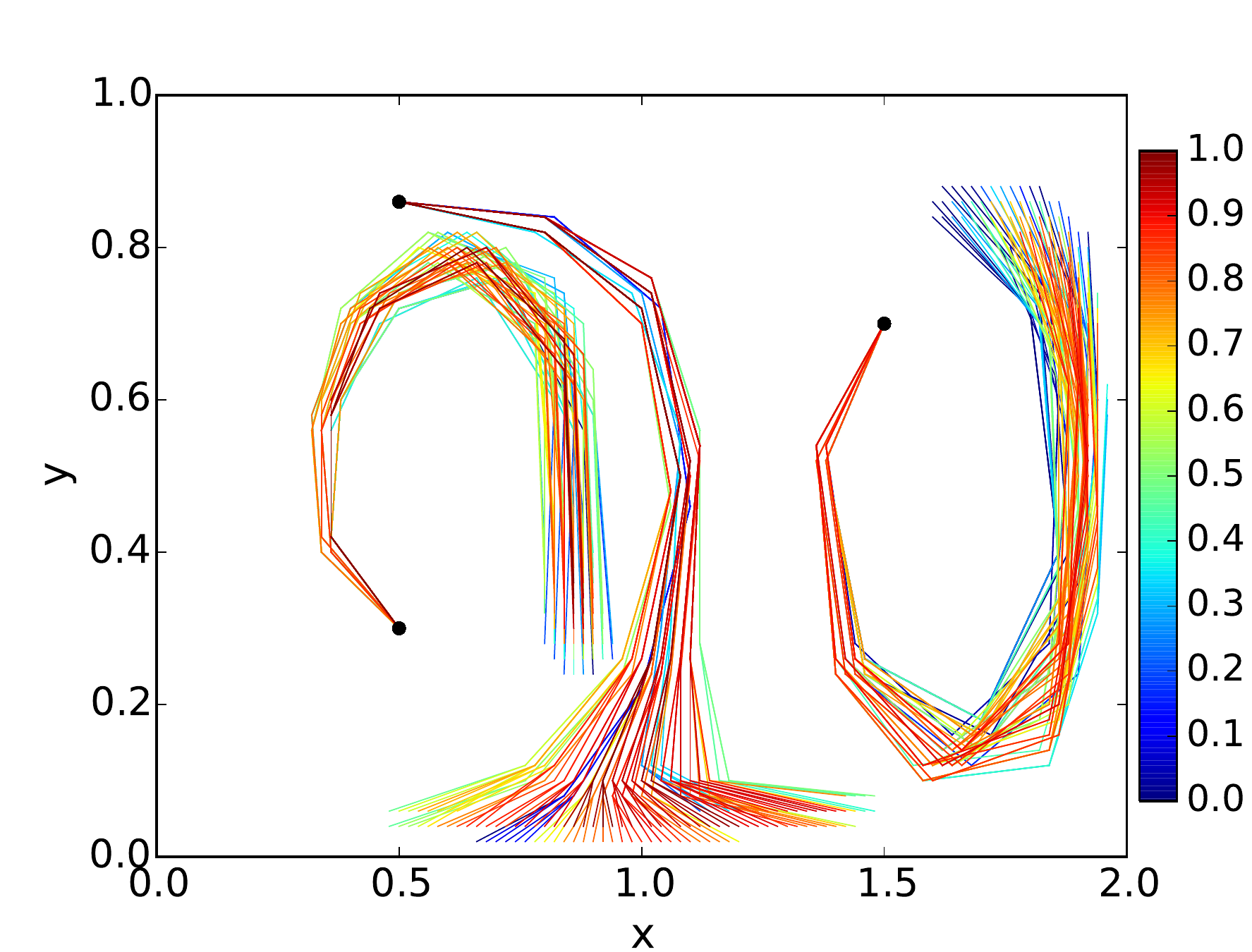}
\caption{6-step paths for the periodically perturbed double gyre at
$\gamma=0.3$, from three different starting nodes (black circles) to
all the accessible destinations. Color coding as in Fig. \ref{fig:A1},
with probability ranges which are $[0.0327,3.355\times10^{-7}]$
(bottom-left node, 66 paths), $[0.0245, 1.879\times 10^{-9}]$ (top-left node,
108 paths), and $[0.0128,1.335\times 10^{-8}]$ (right node, 106 paths).}
\label{fig:A2}
\end{figure}

Figure \ref{fig:A2} shows optimal paths for the periodically
perturbed flow ($\gamma=0.3$). The general clockwise and
counterclockwise rotations still remain, but now there are
pathways connecting the two halves of the domain. Note the
strong divergence of close pathways when they approach the
hyperbolic region at the bottom of the domain, and how is this
geometric structure what allows transport of fluid between the
two regions that were isolated in the steady case.

% figure A3
\begin{figure}
\centering
\includegraphics[width=\columnwidth]{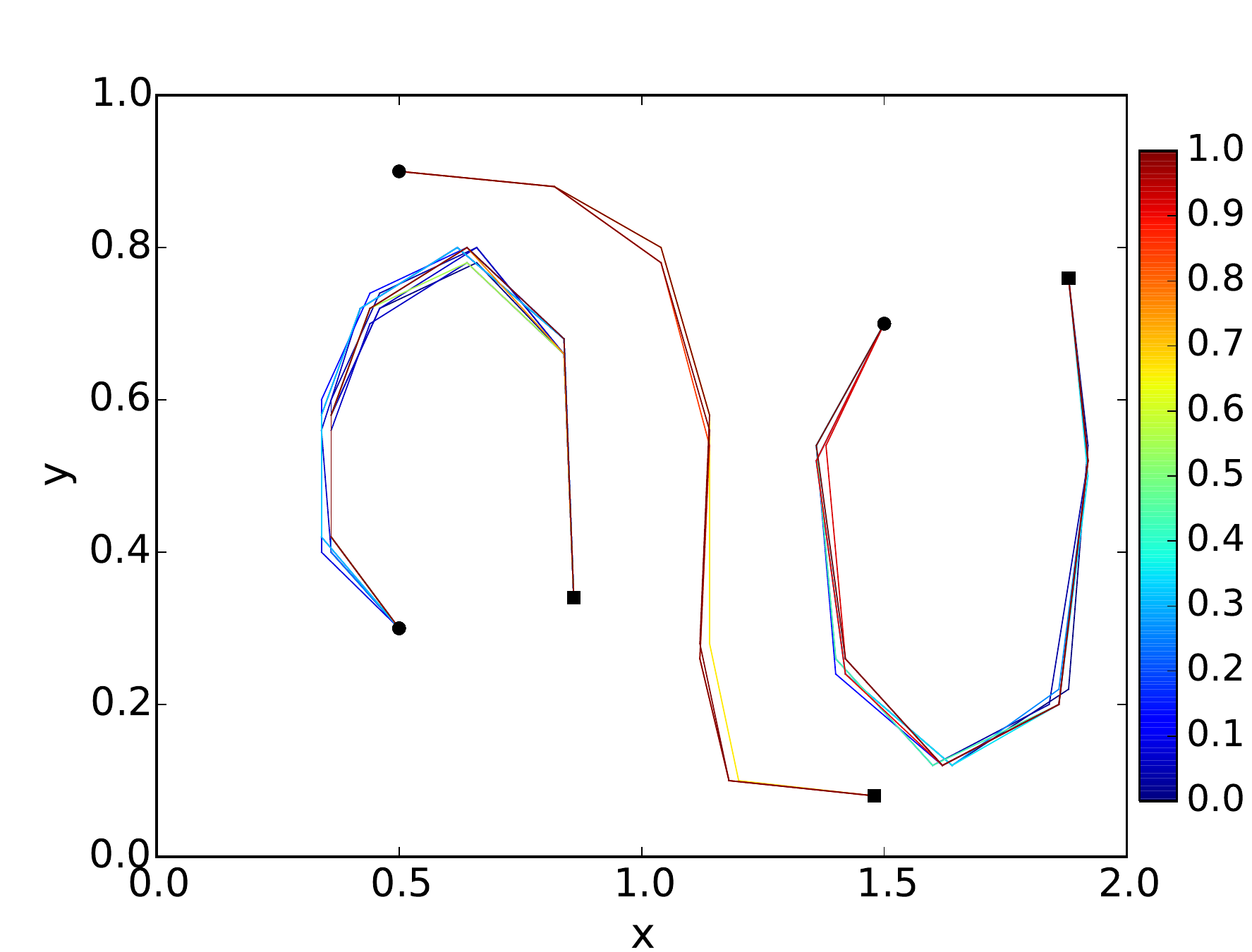}
\caption{Three sets of HPPs in the double gyre for $\gamma=0.3$. We show all HPPs in
$\mathcal{K}_{IJ}^{6}(r=1,\epsilon=0.05)$, with starting nodes $I$ at the black circles
and destinations $J$ at the black squares.
Color coding as in Figs. \ref{fig:A1} and \ref{fig:A2}, with probability ranges
$[0.0282,0.00208]$  (bottom-left node, 12 paths),
$[0.0321,0.00179]$  (top-left node, 7 paths), and
$[0.0128,0.000946]$  (right node, 10 paths). }
\label{fig:A3}
\end{figure}

In Fig. \ref{fig:A3} we display sets of HPPs between three
pairs of nodes at $\gamma=0.3$. More specifically we compute
the paths obtained with $r=1$ and a probability larger than 5\%
of the $P^M_{IJ}$ for these pairs of nodes (i.e. the paths in
the set $\mathcal{K}_{IJ}^{6}(r=1,\epsilon=0.05)$). The HPPs
arrange in very narrow tubes around the MPP, which is the same
behavior observed in the atmospheric paths and also in ocean
calculations \cite{sergiacomi2015most}. The central path in
Fig. \ref{fig:A3} clearly identifies the pathway followed by
particles to connect the left and right regions, using the
``opening" around the hyperbolic region at the top of the
domain.

% Create the reference section using BibTeX:
%\bibliography{references}
%\end{document}

% or copying here the bbl file:

%

\end{document}